\documentclass[a4paper,11pt]{article}

\usepackage{natbib}

\usepackage[dvips]{graphicx}
\usepackage{amsmath}

\begin{document}

\begin{center}
$~$\\
\vspace{1.5 cm}
{\LARGE SEA ICE BRIGHTNESS TEMPERATURE\\AS A FUNCTION OF ICE THICKNESS\\}
\vspace{0.2cm}
{\LARGE Computed curves\\for AMSR-E and SMOS\\
(frequencies from 1.4 to 89 GHz)\\}
\vspace{1.5cm}
{\Large Peter Mills\\}
\vspace{0.4 cm}
Peteysoft Foundation\\
1159 Meadowlane,
Cumberland ON,
K4C 1C3 Canada\\
\vspace{1cm}
{\Large Georg Heygster\\}
\vspace{0.4 cm}
Institute of Environmental Physics, 
University of Bremen\\
Otto-Hahn-Allee 1,
28359 Bremen,
Germany\\
\vspace{0.2cm}
\includegraphics[width=0.75\textwidth]{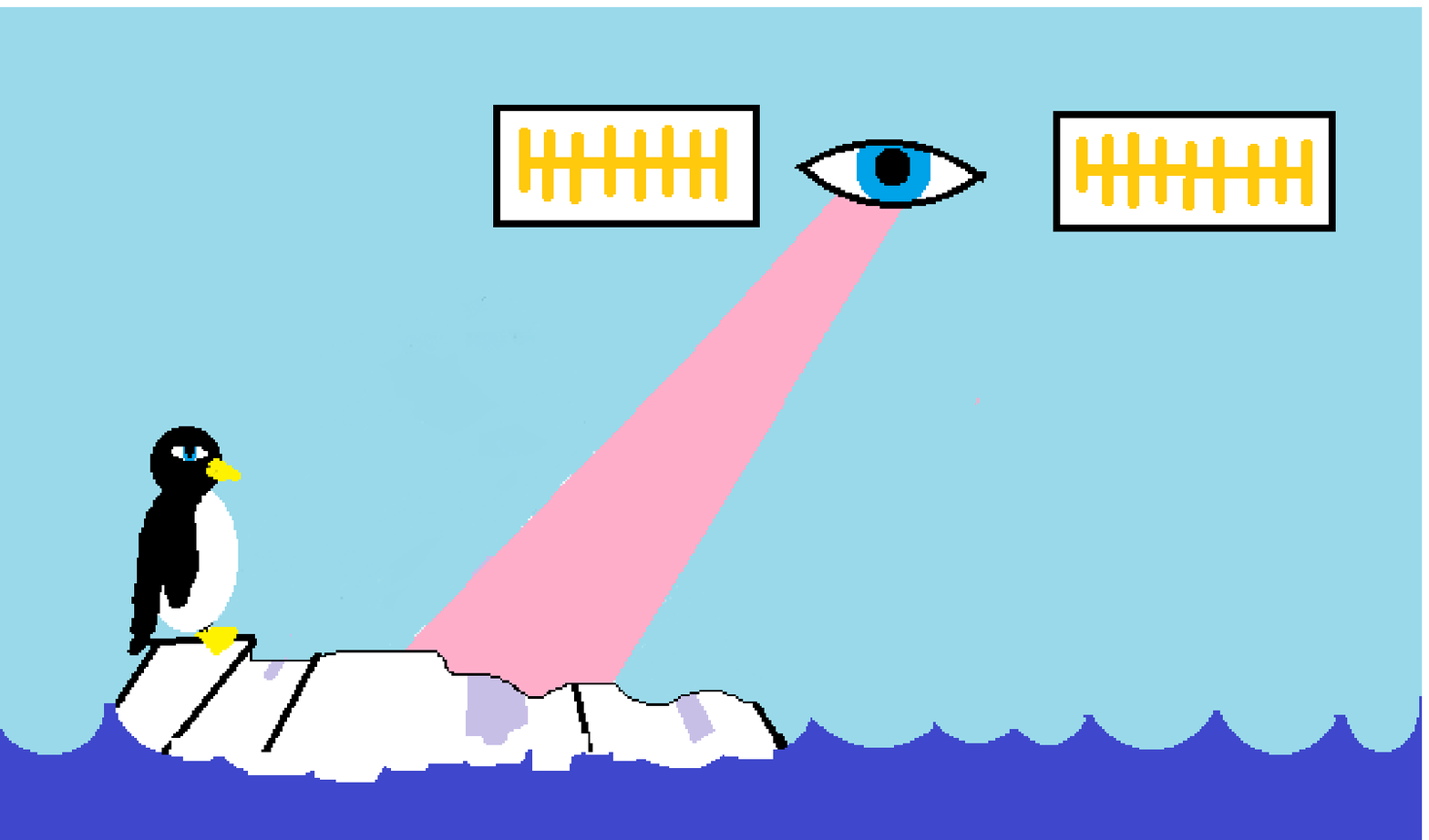}\\
\vspace{0.2cm}
{\large Final report for DFG project HE-1746-15\\
November 16, 2011}

\end{center}
\clearpage

\section{Introduction}

The purpose of this study is to examine the functional dependence of
sea ice brightness temperature on ice thickness 
at frequencies measured by 
the Advanced Microwave Scanning Radiometer on EOS (AMSR-E).
That is, to generate a set of brightness temperature-thickness
($T_b$-$h$) curves.
Several studies have demonstrated a relationship between ice thickness
and SSM/I or AMSR-E brightness temperatures \citep{Martin_etal2004,
Naoki_etal2008,Kwok_etal2007,Hwang_etal2007} with at least one attempt to use
this relationship for the purpose of ice thickness retrieval \citep{Martin_etal2004}.

This relationship, however, is not a direct one in the sense that,
all other things begin equal, changes in ice thickness will not necessarily affect 
microwave emissions at AMSR-E frequencies.  Even at 6.8 GHz,
the penetration depth is far smaller than all but the thinnest of ice sheets.
The relationship is caused by the fact that thinner, newer ice, has different
physical properties than those of older, thicker ice.  In particular
it tends to be more saline, especially in the top layers.

Three processes ensure that newer, thinner ice is, on average, more saline
than older, thicker ice 
\citep{Tucker_etal1992,Eicken1992,Weeks_Ackley1985,Vancoppenolle_etal2007,Ehn_etal2007}, 
especially in the top layers.
As new ice is formed, most of the salt gets expelled, except for a small amount
that is included as pockets of highly saline brine.  The faster the
ice grows, the more brine is included.  Since thin ice conducts heat more
quickly, it will grow faster than thicker ice on average.
As a growing ice sheet cools, some of the water in the brine will freeze.
Since ice is less dense than brine, increasing pressure in the brine pockets will
cause some of the brine to be expelled from both the top and bottom
of the ice sheet, producing the characteristic 'C'-shaped profile of
first-year ice.  
Finally, as the ice ages, thawing will produce channels through
which much of the brine can drain.

The causal relationship between salinity and brightness temperature is similarly
circuitous, although it is a more definite one.  Higher salinity produces a higher
effective permittivity whose main effect, for optically thick ice, 
is to increase the polarization difference or second Stokes parameter.
It is not just the salinity that affects the microwave signal.  Changes in
temperature will have a similar effect as well as a similar cause.  That is, warmer
ice will tend to have a higher effective permittivity and thinner ice will tend to
be warmer, again because of heat conduction.

We use parameterised empirical relationships
between ice thickness and salinity.  Model profiles are fed to ice emissivity models
to determine idealized relationships between brightness temperature and ice thickness.

\section{Model set up}

\begin{figure}
\includegraphics[width=0.9\textwidth]{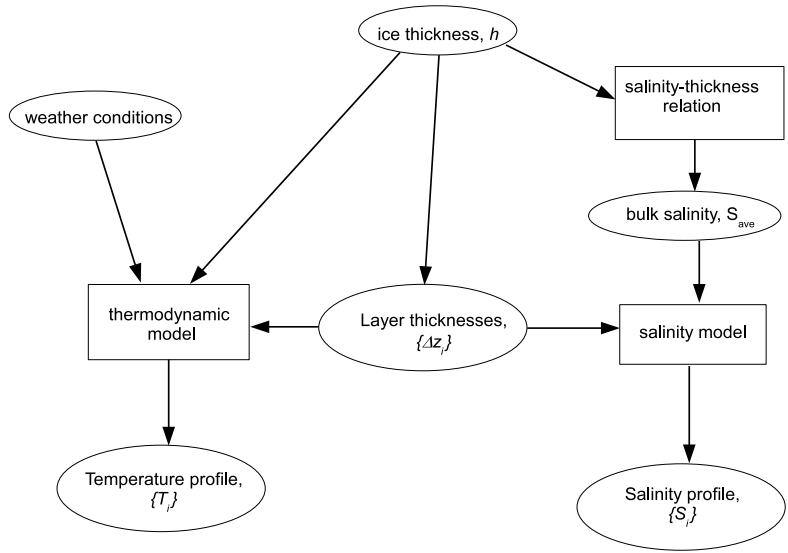}
\caption{Data flow diagram describing the model set up, namely how the other parameters
are derived from ice thickness.  Rectangles are processes, ellipses are data.
The salinity-thickness relation and the salinity model are described in 
Section \ref{salinity_section}.  
The thermodynamic model is described in Section \ref{thermo_section}}
\label{setup_flow}
\end{figure}

\begin{figure}
\includegraphics[width=0.9\textwidth]{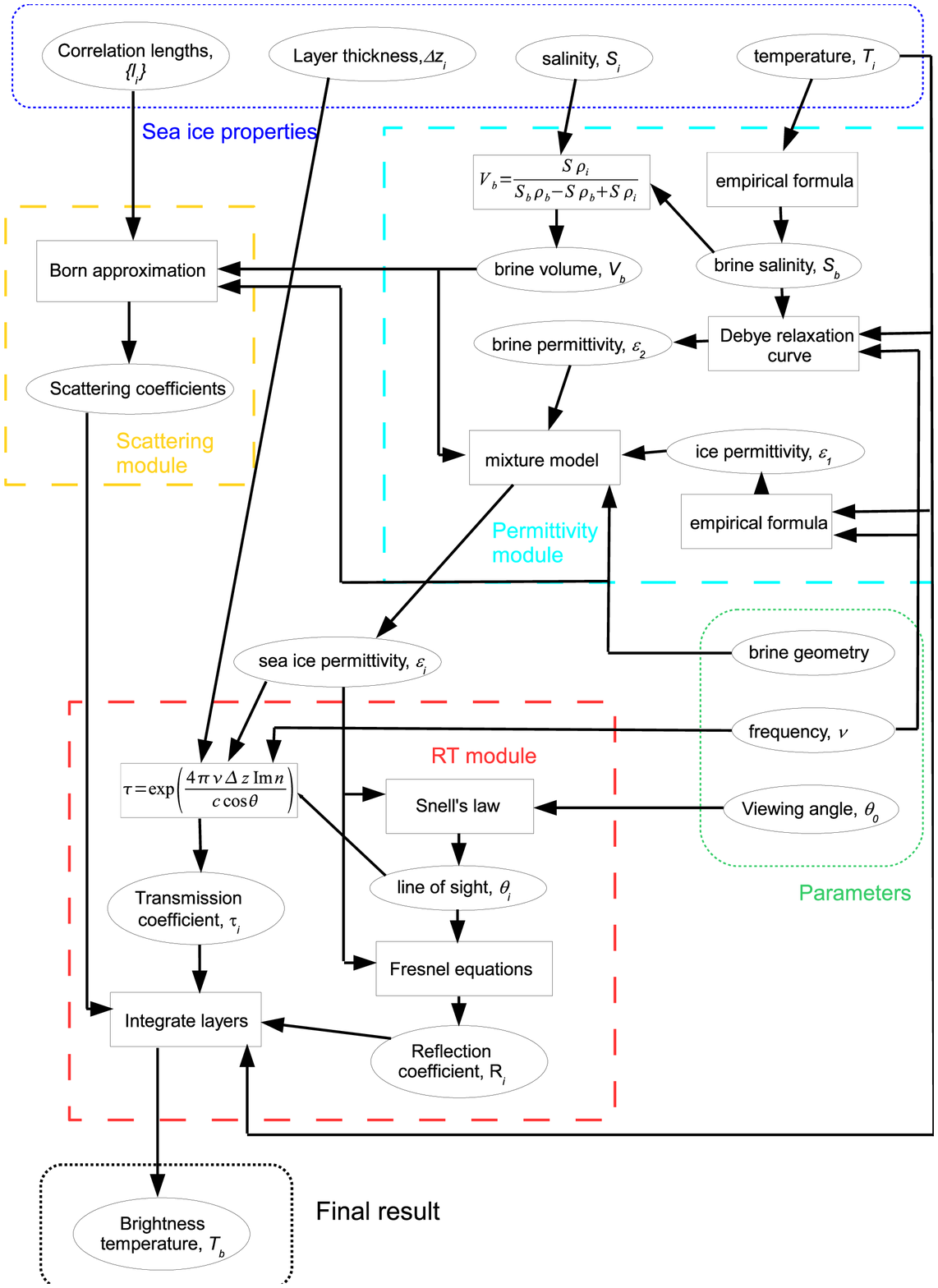}
\caption{Detailed data flow diagram describing the complete ice emissivity model.
Rectangles are processes, ellipses are data.
The ``mixture model'' is described in Section \ref{eps_section}.
Other components are described elsewhere, see for instance 
\citet{Mills_Heygster2011a} and \citet{Wiesmann_Maetzler1999a}.}
The boxes labelled, respectively,
 ``Snell's law'', ``Fresnel equations'', ``Integrate layers''
and the box with the equation for the transmission coefficent 
correspond to equations (3), (1-2), (5-6) and (7-8) in
\citet{Mills_Heygster2011a}.
The equations for fresh-water ice permittivity are contained
in \citet{Hufford1991}.
For other equations in the permittivity module, 
a good reference is \citet{Ulaby_etal1986}.
\label{data_flow}
\end{figure}

We use a layered, plane-parallel, radiative transfer (RT) model called Microwave
Emissivity Model for Layered Snowpack (MEMLS)
\citep{Wiesmann_Maetzler1999a,Wiesmann_Maetzler1999}.  
The emissivity calculation can be divided into three main modules.
First, effective relative permittivities must be calculated from the physical
properties of each layer--chiefly, temperature and salinity.  
Second, scattering coefficients must also be calculated
from values of correlation length.  
Finally, these permittivities and scattering coefficients
are simultaneously integrated through the depth of the ice sheet.
Before we can feed these quantities into the model, however,
we must take account of the numerous and complex 
interrelationships between the different inputs,
particularly since we are interested only in the functional
dependence of brightness temperature on a single bulk property,
i.e. ice thickness.
For example, because of changes in relative brine volume
at different temperatures and salinities, 
correlation lengths of the brine inclusions
will be affected by these other two properties.

The emissivity model is described in the data flow diagram
shown in Figure \ref{data_flow} including data inputs
and parameters.
The basic modules have a dashed outline.
As can be seen, it is a quite a complex model comprising 
many separate yet interacting components.
The model set up is shown in Figure \ref{setup_flow}.

Both ice thickness and the salinity profile (function of salinity with
depth) will be affected by past weather conditions.
In addition, the current temperature profile will be
determined by both the ice thickness and prevailing weather
conditions.  For this reason, we have designed a simple 
thermodynamic model which can be used both to determine
the temperature profile, and in a crude fashion, model
ice growth.  This model is described below, along with the
other components.

The calculation of effective permittivities is described below.  For scattering 
coefficients, we use the empirical model within MEMLS \citep{Wiesmann_Maetzler1999}.
We relate the correlation length to brine volume as follows:
\begin{equation}
l = l_0 \sqrt{V_b}
\end{equation}
where $l_0$ is a constant and $V_b$ is relative brine volume.
For the case of randomly positioned, oriented cylinders, the correlation length
in the transverse direction would be related to brine volume
to the first power.
For the case of randomly positioned spheres, it would be to the third power.
\citep{Maetzler1997}
We take the second power, thus the brine inclusions are somewhere
between a cylinder and a sphere.

The radiative transfer model will not be described in this report
since it and similar models are covered in other publications
--- see for instance, \citet{Wiesmann_Maetzler1999a}
and \citet{Mills_Heygster2011a}.
We will note, however, that there is some descrepancy in the
two models contained in the aforementioned publications as to
when and how (by simply eliminating it or taking the absolute value)
to drop the imaginary component of the reflection coefficients.
In MEMLS, for instance, only the real part of the complex
permittivity is used in the calculation of reflection coefficients.
In \citet{Mills_Heygster2011a}, by contrast, the imaginary
part is carried right to the end of the calculation and 
real reflection coefficients produced by taking the absolute
value.
Using one method or the other can mean a difference of as much
as 10 K.
We have modified the MEMLS code so that calculations of reflection
coefficients match the model described in \citet{Mills_Heygster2011a}.
When scattering is neglected in MEMLS, the two models agree to within
half a Kelvin.

The viewing angle is set at 55 degrees, or the same
as AMSR-E.
 
\subsection{Model for effective permittivities}

\label{eps_section}

\begin{figure}
\includegraphics[width=0.9\textwidth]{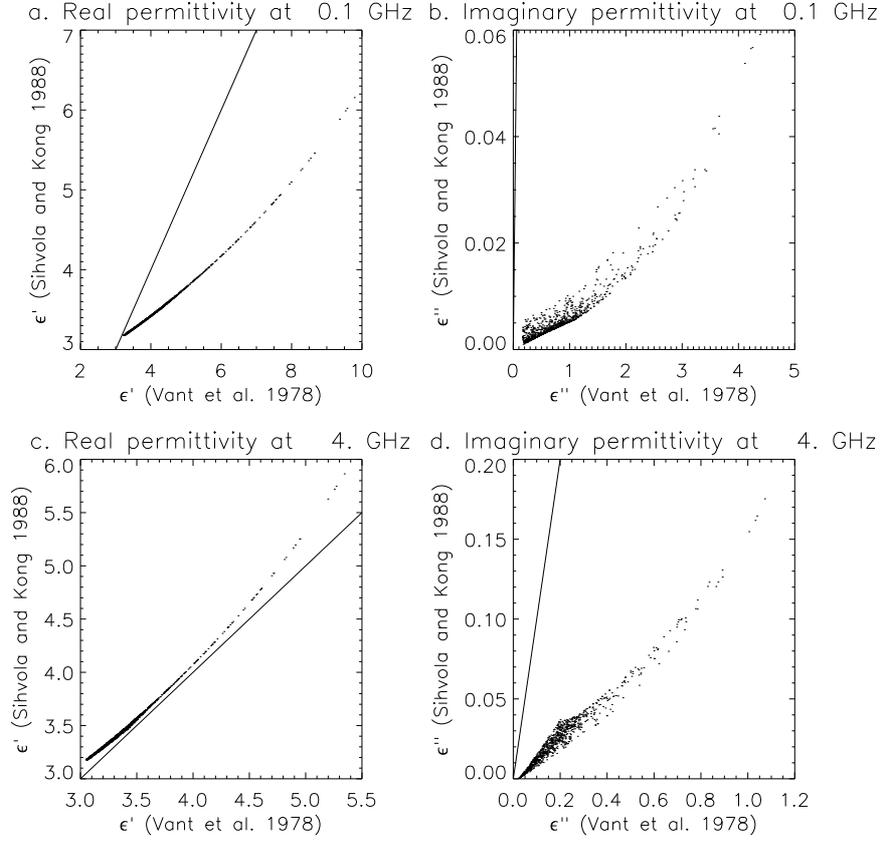}
\caption{Comparison of the mixture model of Sihvola and Kong based on de-polarizability
and the empirical mixture model of Vant using random values of temperature
in the range of 251 to 271 K and salinity in the range of 0 to 15 psu.}
\label{mix_compare}
\end{figure}

\begin{table}
\caption{Correlation between Vant and Sihvola mixture
models of effective permittivity for both real and
imaginary components.}
\label{eps_corr}
\begin{center}
\begin{tabular}{|l|c|c|}\hline
$\nu$ & $r (\epsilon^\prime)$ & $r (\epsilon^{\prime\prime})$ \\
\hline\hline
  0.1 &  0.994&  0.929\\
  0.2 &  0.994&  0.928\\
  0.4 &  0.995&  0.942\\
  0.8 &  0.995&  0.974\\
  1.0 &  0.996&  0.981\\
  2.0 &  0.995&  0.984\\
  4.0 &  0.995&  0.983\\

\hline
\end{tabular}
\end{center}
\end{table}

\begin{figure}
\includegraphics[width=0.9\textwidth]{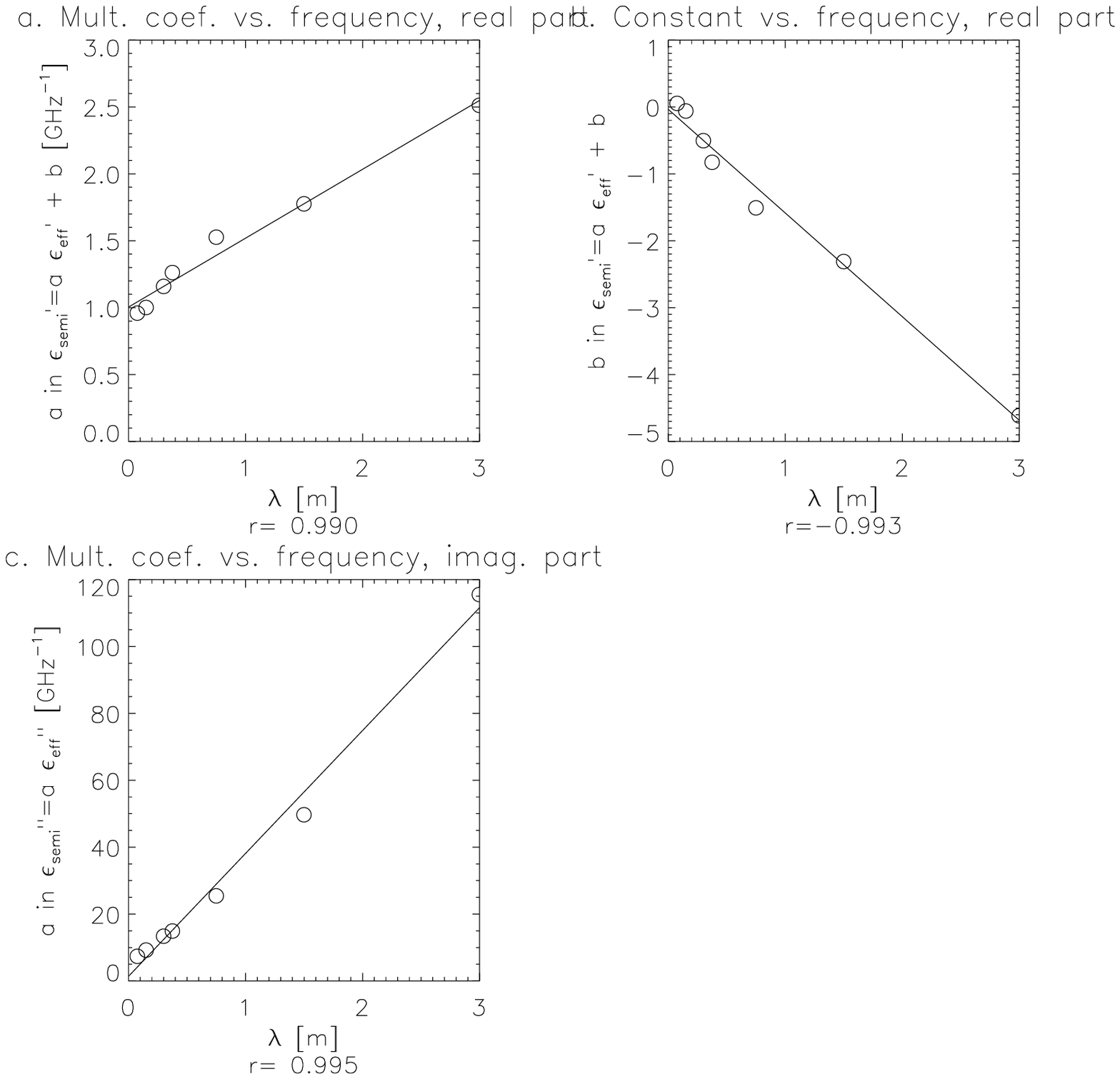}
\caption{Correlations with wavelength of
 regression coefficients between empirical and theoretical
mixture models.}
\label{coeff_vs_nu}
\end{figure}

\begin{figure}
\includegraphics[width=0.45\textwidth]{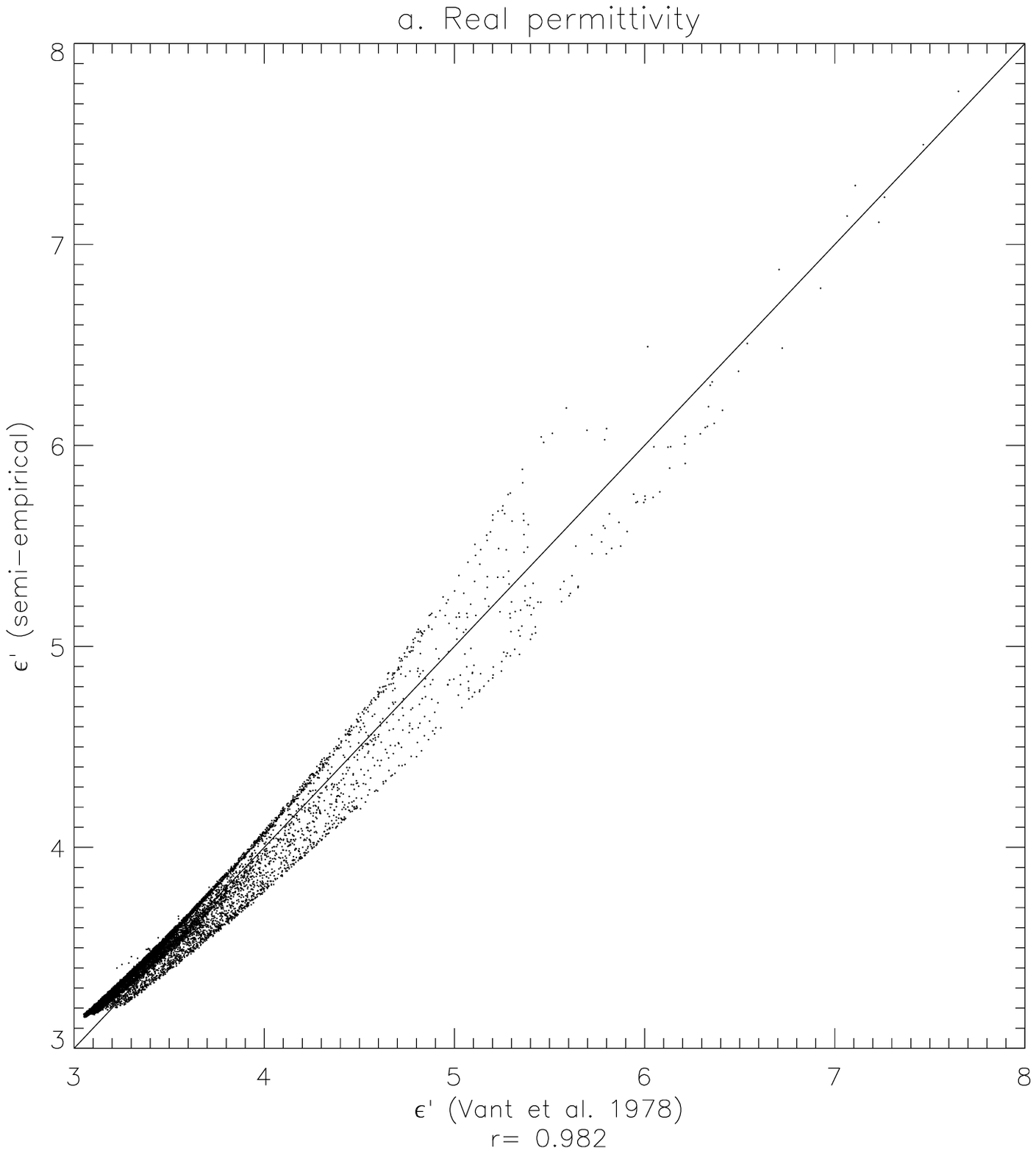}
\includegraphics[width=0.45\textwidth]{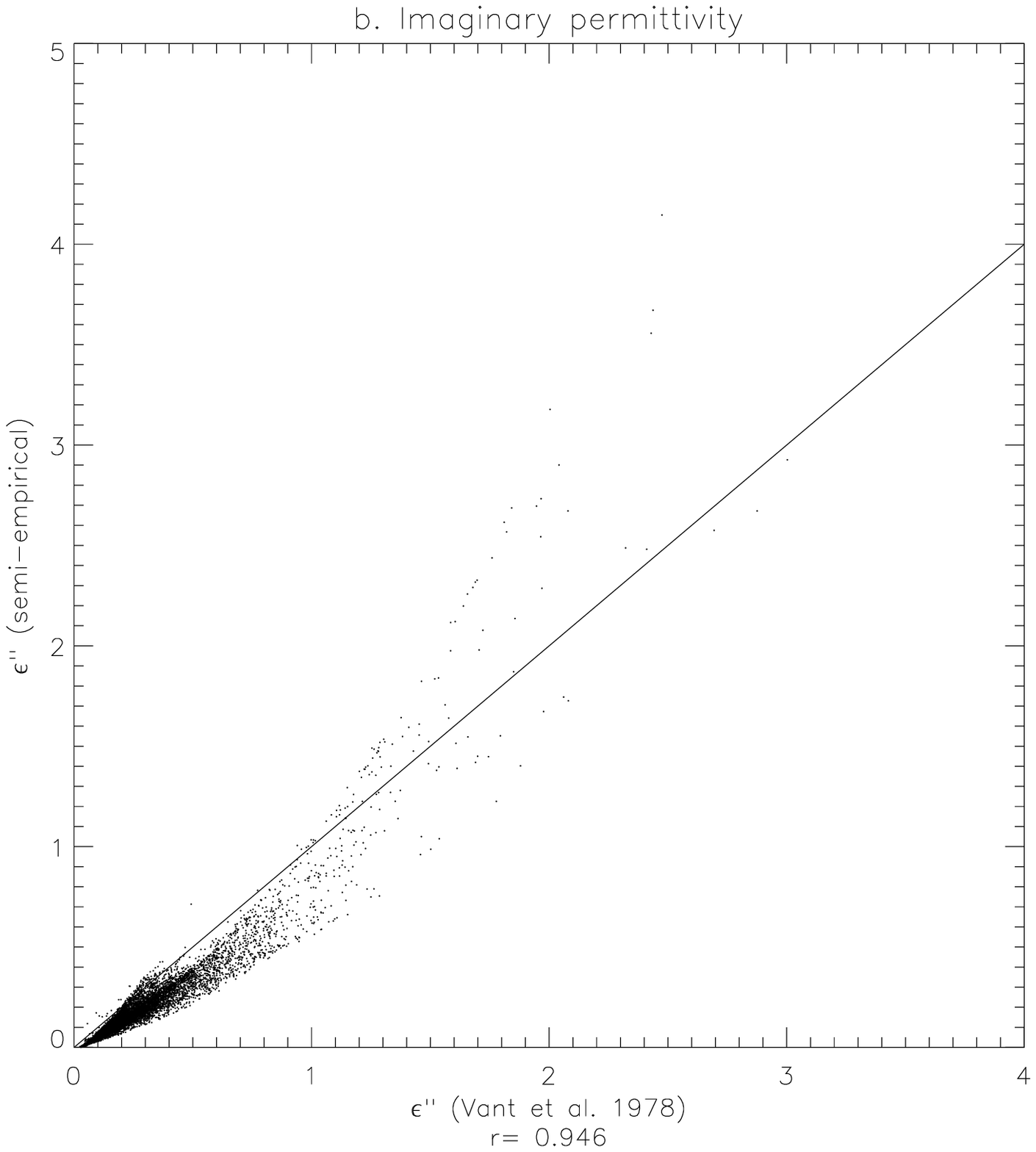}
\caption{Validation of semi-emprical mixture model.
	All frequencies between 0.1 and 4 GHz.}
\label{semi_compare}
\end{figure}

\begin{table}
\caption{Fitted coefficients for semi-empirical mixture model.}
\label{eps_semi_coeff}
\begin{center}
\begin{tabular}{|l|r|}
\hline
Coefficient & value \\
\hline\hline
$c_1$ &     0.154\\
$c_2$ &     1.005\\
$c_3$ &    -0.464\\
$c_4$ &    -0.036\\
$c_5$ &    11.011\\
$c_6$ &     1.409\\

\hline
\end{tabular}
\end{center}
\end{table}

The calculation of effective relative permittivity is one of the most
important steps in the model.
It is also the one of the least understood and most fraught with
uncertainty.
To calculate the effective permittivity,
we use a semi-empirical model combining the empirical mixture model of 
\citet{Vant_etal1978} and the theoretical one from \citet{Sihvola_Kong1988}
based on the low-frequency limit.
Vant et al. give an empirical mixture model based on fits of
measured real and imaginary permittivity to
brine volume:
\begin{equation}
\epsilon^* = a V_b + b
\label{Vant}
\end{equation}
where $\epsilon^*$ is the effective permittivity (real or imaginary),
$V_b$ is the relative brine volume and $a$ and $b$ are constants.
Different constants have been acquired for frequencies between
0.1 and 4 GHz.
To extend the frequency range of the Vant models, we combine it with the
following theoretical, mixture model for oriented needles based on the low-frequency
limit
\citep{Sihvola_Kong1988}:
\begin{equation}
\epsilon_{eff}=\epsilon_1 + \frac{V_b (\epsilon_2 - \epsilon_1) \epsilon_1 /
		\left [\epsilon_1 + P (\epsilon_2 - \epsilon_1)\right ]}
		{1 - P V_b (\epsilon_2 - \epsilon_1)\left [\epsilon_1 + P (\epsilon_2 - \epsilon_1) \right ]}
\label{mixture}
\end{equation}
where $\epsilon_1$ is the permittivity of the host material (ice), 
$\epsilon_2$ is the permittivity of the inclusion material (brine),
and $P=0.5$ is the de-polarisation factor.
It was observed that the results from (\ref{Vant}) and (\ref{mixture})
are closely correlated as shown in Figure \ref{mix_compare}, 
however they differ by both a constant factor
(in the real part) and a coefficient 
(in both the real and imaginary parts).
These parameters were found to vary closely with the 
wavelength, as shown in Figures \ref{coeff_vs_nu}.

Thus, the combined model is given as follows:
\begin{equation}
\epsilon_{semi} = (c_1/\nu+c_2) \epsilon_{eff}^\prime + c_3/\nu+c_4 +
		(c_5/\nu + c_6) \epsilon_{eff}^{\prime\prime} i
\end{equation}
where $\epsilon_{eff}=\epsilon_{eff}^\prime+\epsilon_{eff}^{\prime\prime} i$
is the effective permittivity from Equation (\ref{mixture}),
$\nu$ is frequency and $c_1$ through $c_6$ are fitted constants.
Note that there is no constant term for the imaginary part,
forcing the value to zero for zero brine volumes 
(pure ice is almost a perfect dielectric).
We compare the derived semi-emprical model with the Vant model
for a range of frequencies between
0.1 and 4. GHz, a range of salinities between 0 and 15 psu
and a range temperatures between 251 and 271 K in Figure \ref{semi_compare}.
The fitted coefficients, $c_1$ through $c_6$
are given in Table \ref{eps_semi_coeff}.

Finally, a note on brine volume.  In \cite{Mills_Heygster2011a},
brine volume was given as:
\begin{equation}
V_b = \frac{S}{S_b}
\end{equation}
where $S_b$ is the brine salinity which, because of freezing
point depression, can be calculated from temperature only.  
This is a rather crude approximation.
A more accurate formula is as follows:
\begin{equation}
V_b = \frac{S \rho_i}{S_b \rho_b - S \rho_b + S\rho_i}
\end{equation}
where $\rho_i$ and $\rho_b$ are the densities of ice and
brine, respectively.  This formula is found by writing
the ratio of brine volume to ice volume in terms of density
and salinity and solving.
It directly connects the calculation of brine salinity
(an accurate function of temperature only) with relative brine volume.

\subsection{Salinity model}

\label{salinity_section}

\begin{figure}
\begin{center}
\includegraphics[angle=90,width=0.9\textwidth]{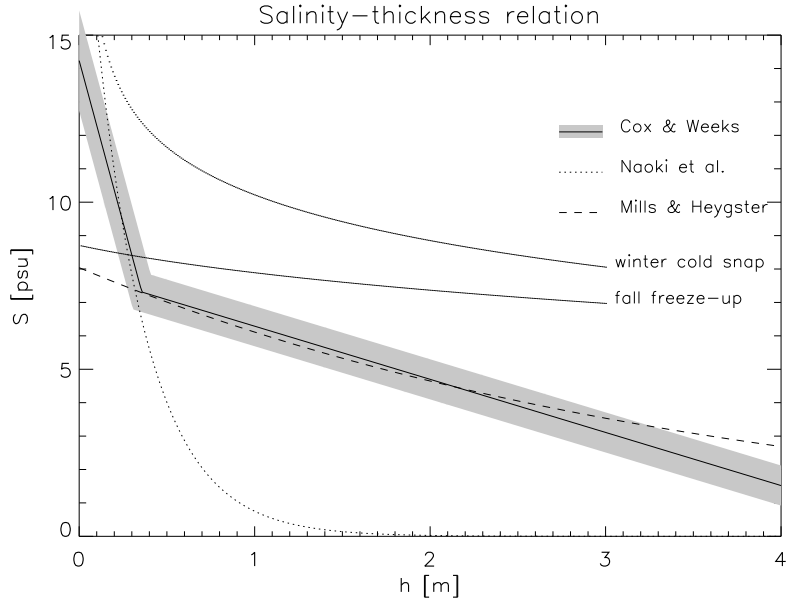}
\caption{Different salinity-thickness relationships compared.
Shading are residuals for the Cox and Weeks models.}
\label{sh_rel}
\end{center}
\end{figure}

\begin{figure}
\begin{center}
\includegraphics[width=0.8\textwidth]{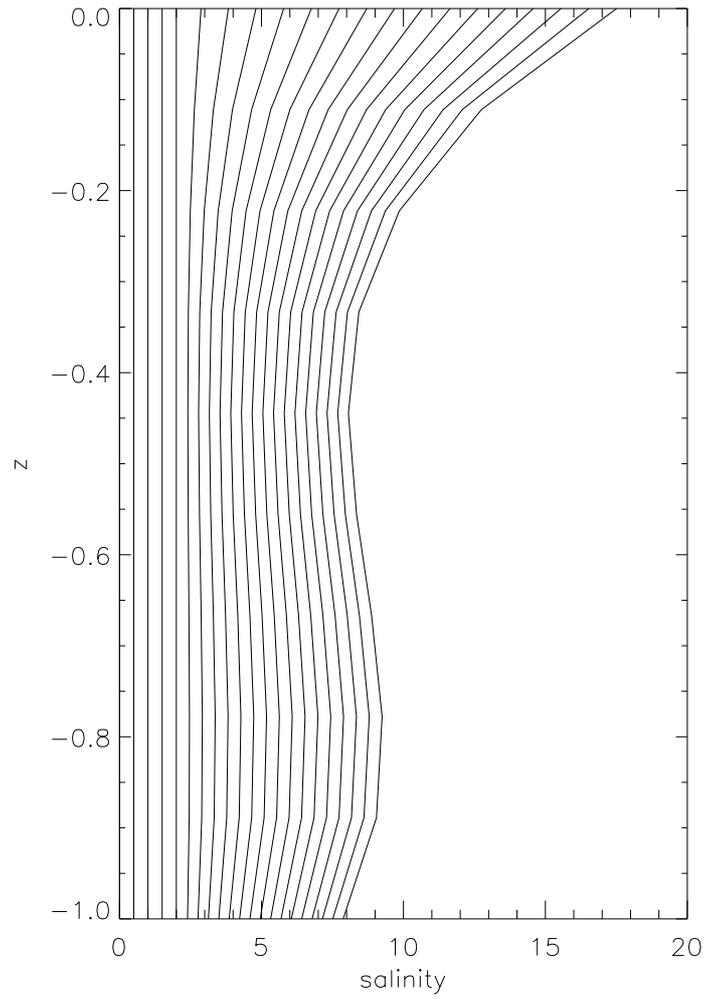}
\caption{Model salinity profiles for different bulk salinities in 0.5 psu increments,
starting at 0.5psu and ending at 10psu.  Ice depth is normalized.}
\label{salinity_profile}
\end{center}
\end{figure}

Figure \ref{sh_rel} shows plots of different salinity-thickness
(S-$h$) curves which have been derived empirically and from a
simple model (see below).
\citet{Cox_Weeks1974} provide the following
piece-wise, linear fit of bulk salinity
to ice thickness (solid curve):
\begin{equation}
S_{\mathrm{ave}} = \left \lbrace 
\begin{array}{lr}
14.24-19.39 h; & h \le 0.4 m \\
7.88-1.59 h; & h > 0.4 m
\end{array} \right .
\end{equation}
where $S_{\mathrm{ave}}$ is bulk salinity and $h$ is ice thickness 
in metres.
This is the only S-h curve that will be used in the study.
Rather than trying the emissivity models with different curves,
\citet{Cox_Weeks1974} provide residuals for the curves.
We will use these residuals, along with those 
supplied with the Vant effective permittivity models to compute an error bound
for the final results (see below).

The other five curves are as follows.
\citet{Naoki_etal2008} generated an exponential fit
(short-dashed curve) of surface salinity (top 10 cm)
based on measurements conducted in the Sea of Okhotsk:
\begin{equation}
S = \exp(m h + k)
\label{exp_fit}
\end{equation}
where $S$ is the salinity, $h$ is ice thickness and $m$ and $k$
are constants.
\citet{SMOSIce_report} generated
an exponential fit of ice core samples
taken from the Weddell Sea \citep{Eicken1992}.
Coefficients for equation (\ref{exp_fit}) are $m=-3.55$
and $k=3.06358$ for the Naoki model and $m=-0.275$ and
$k=8.04$ for \citet{SMOSIce_report}.
Notice that the Naoki curve roughly matches the left section
of the Cox and Weeks curve, while the Mills and Heygster
curve roughly matches the right section.

The dotted curves show results from the thermodynamic ice growth
model (described in the next section) for two weather scenarios:
fall freeze-up and a winter cold snap.
These provide a reasonable range for thin ice (below 30 cm), however
they over-estimate salinity for thick ice.

Two trials will be performed:  one in which the ice is represented as a single layer,
and one with a parameterized salinity profile.  For the parameterized salinity profile,
we use the 'S'-shaped model from \citet{Eicken1992}, but with the caveat that at values
for bulk salinity below 2psu, it reverts to a flat profile in accordance with
\citet{Granskog_etal2006}.  
This is shown in Figure \ref{salinity_profile}.

\subsection{Thermodynamic model}

\label{thermo_section}

\begin{table}
\caption{Thermodynamic model parameters for the two weather scenarios}
\label{thermo_parm}
\begin{center}
\begin{tabular}{|l|cc|}
\hline
 & Scenario & \\
\hline
Parameter & Fall freeze-up & Winter cold snap \\
\hline\hline
Wind speed [m/s] & 2. & 10. \\
Air temperature [K] & 270. & 260. \\
Relative humidity & 0.5 & 0.1 \\
Cloud-cover & 0.5 & 0.1 \\
Insolation [W/m$^2$] & 50. & 0. \\
\hline
\end{tabular}
\end{center}
\end{table}

As mentioned in the introduction, it is not just ice salinity that
lowers the brightness temperature of thin ice, it is also the temperature:
thin ice conducts heat faster.  This increase in temperature will
have two effects: direct (hotter objects are brighter) and indirect
(higher temperatures produce higher complex permittivities.)
Therefore, for the multi-layer models, we calculate the surface
temperature using a thermodynamic model based on certain assumed
prevailing weather conditions.  Since we assume the ice is in
thermal equilibrium---valid if the ice is not too thick
and the weather is changing relativly slowly---
the temperature profile is linear.

The following equation relates the ice surface temperature to
net heat flux:
\begin{equation}
h Q^* = k (T_w - T_s)
\label{thermo_model}
\end{equation}
where $T_w$ is the water temperature which is assumed to be
constant at freezing (approximately -1.9$^\circ$ C at a water
salinity of 35 psu), $T_s$ is surface temperature and $k$ is 
the thermal conductivity of the ice.
The net heat flux comprises the following components,
with functional dependencies supplied:
\begin{equation}
Q^* = Q_E [e(T_s)] + Q_H (T_s) + Q_{SW} (T_s^4) + Q_{LW}
\label{net_heat}
\end{equation}
The terms on the RHS are, from left to right:
latent heat, sensible heat, longwave and shortwave;
$e(T)$ is the saturation vapour pressure.
The first two terms are approximated with simple
parameterisations while the longwave flux is based
on the Stefan-Boltzmann law.
The shortwave flux is calculated primarily from
geometric considerations based on the position
of the Earth relative to the Sun.
The following inputs are required for the model:
surface- wind speed, humidity, air temperature
and density (or pressure), cloud cover and
date and time or insolation.
\citep{Cox_Weeks1988,Drucker_etal2003,Yu_Lindsay2003}
Inputs can be supplied to give a 
picture of the general weather conditions.
For instance, fall freeze-up might be characterized
by relatively mild temperatures, low winds, high humidity,
high cloud cover and moderate insolation.
By contrast, a winter cold snap would be characterized by low
temperature, high winds, low humidity, clear conditions
and little to no insolation. See Table \ref{thermo_parm}.
Equations (\ref{thermo_model}) and (\ref{net_heat})
are solved with a numerical root-finding algorithm,
specifically bisection \citep{nr_inc2}.

The model can also be used as a crude ice growth model.
The rate of ice growth is:
\begin{equation}
g = \frac{Q^*}{L \rho_i}
\end{equation}
where $L$ is the latent heat of fusion for water.
Empirical equations for determining the initial brine entrapment 
in sea ice have been derived by 
\citet{Cox_Weeks1988} and \citet{Nakawo_Sinha1981} and take the form:
\begin{equation}
S = S_0 f(g)
\end{equation}
where $S_0$ is the salinity of the parent water and $f$ is
an empirical function of ice growth rate.

As a qualitative validation of the salinity-thickness
relations described in \ref{salinity_section},
this model is quite effective.
As seen in Figure \ref{sh_rel}, the model considerably overestimates
the actual salinities, especially as the ice get thicker.
This is because it does not take into account brine drainage and
expulsion.

\subsection{Error}

\begin{figure}
\includegraphics[width=0.45\textwidth]{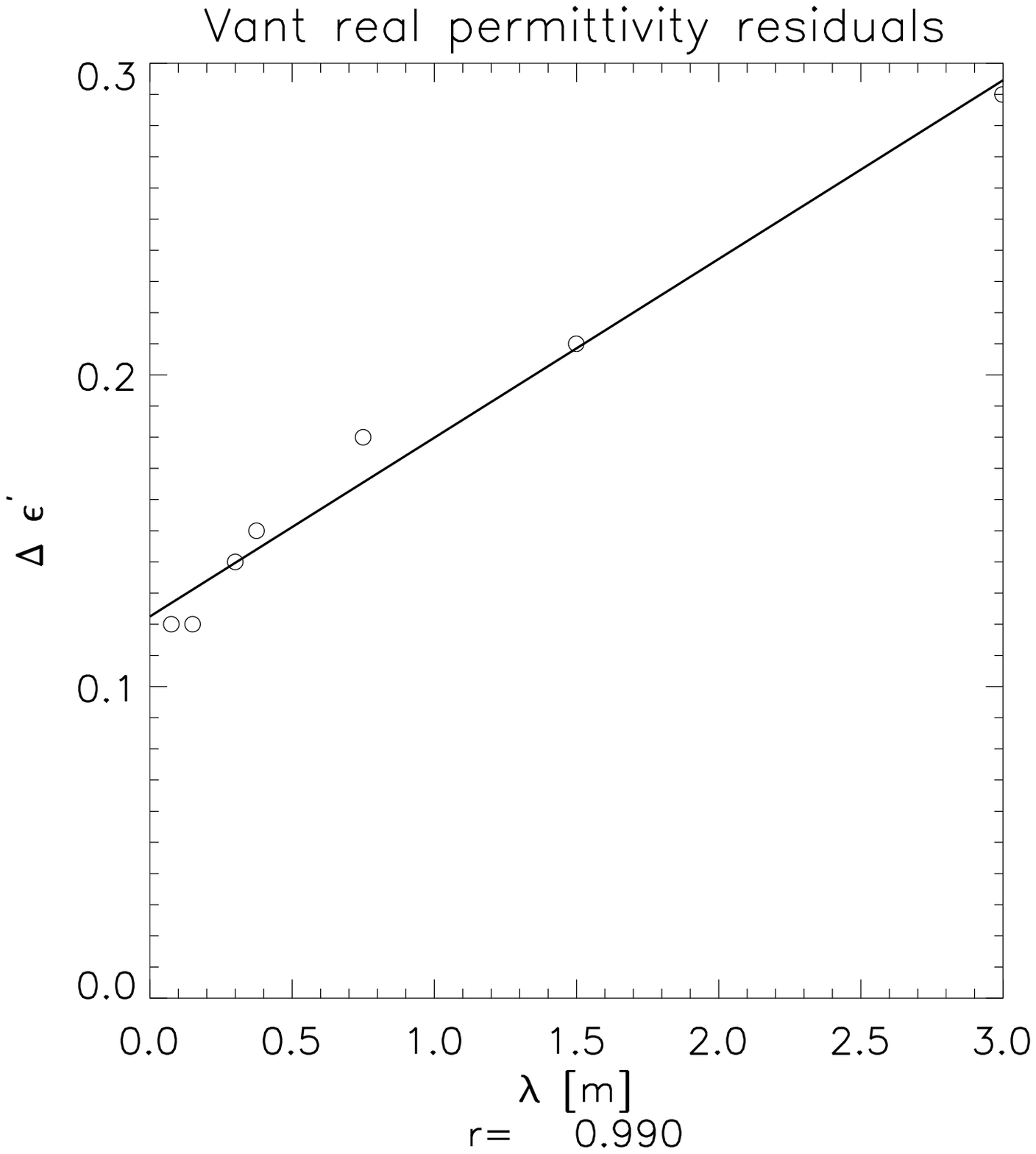}
\includegraphics[width=0.45\textwidth]{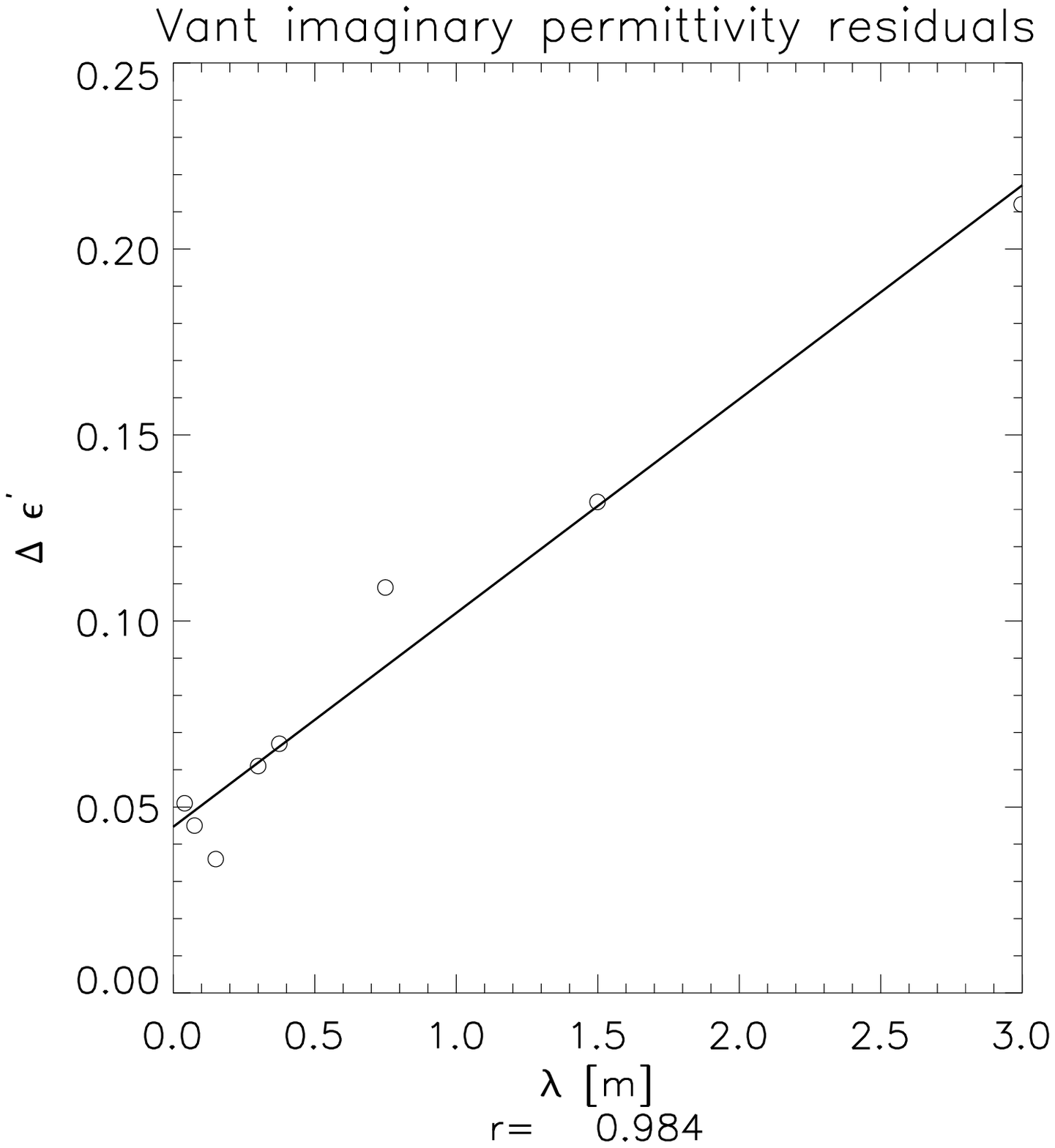}
\caption{Error residuals for the Vant effective permittivity models
	plotted against wavelength.  Fitted curves are shown.}
\label{err_vs_nu}
\end{figure}

Both the salinity-thickness relation from \citet{Cox_Weeks1974} and the mixture models from
\citet{Vant_etal1978} are supplied with statistical error bounds
(root-mean-square errors (RMSE) or residuals).
We use these to generate error tolerances for the final $T_b$-thickness curves.
Residuals from the Vant models were found to have a similar
close correlation with wavelength as the relationship
between the two effective permittivity models:
see Figure \ref{err_vs_nu}.
This relationship was use to extrapolate the residuals
to higher frequencies.
A Monte Carlo or ``bootstrapping'' \citep{nr_inc2} method is used.  
Intitial salinities are perturbed
with Gaussian deviates with zero mean and
standard deviation equal to the residuals.  So are the
generated complex permittivities.
The number of trials was chosen based on convergence of error estimates.

\section{Results}

\begin{figure}
\begin{center}
\includegraphics[angle=90,width=0.9\textwidth]{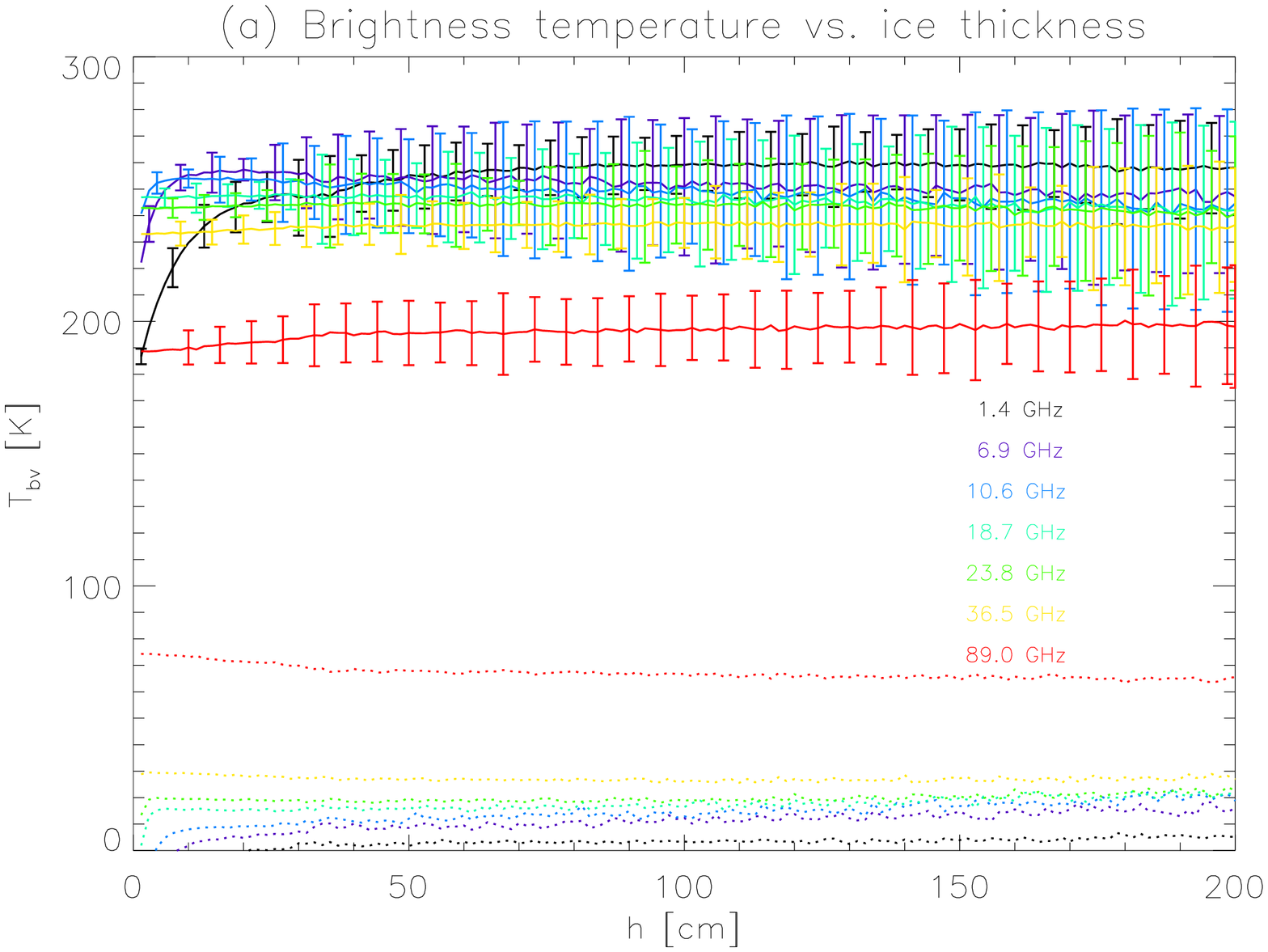}
\includegraphics[angle=90,width=0.9\textwidth]{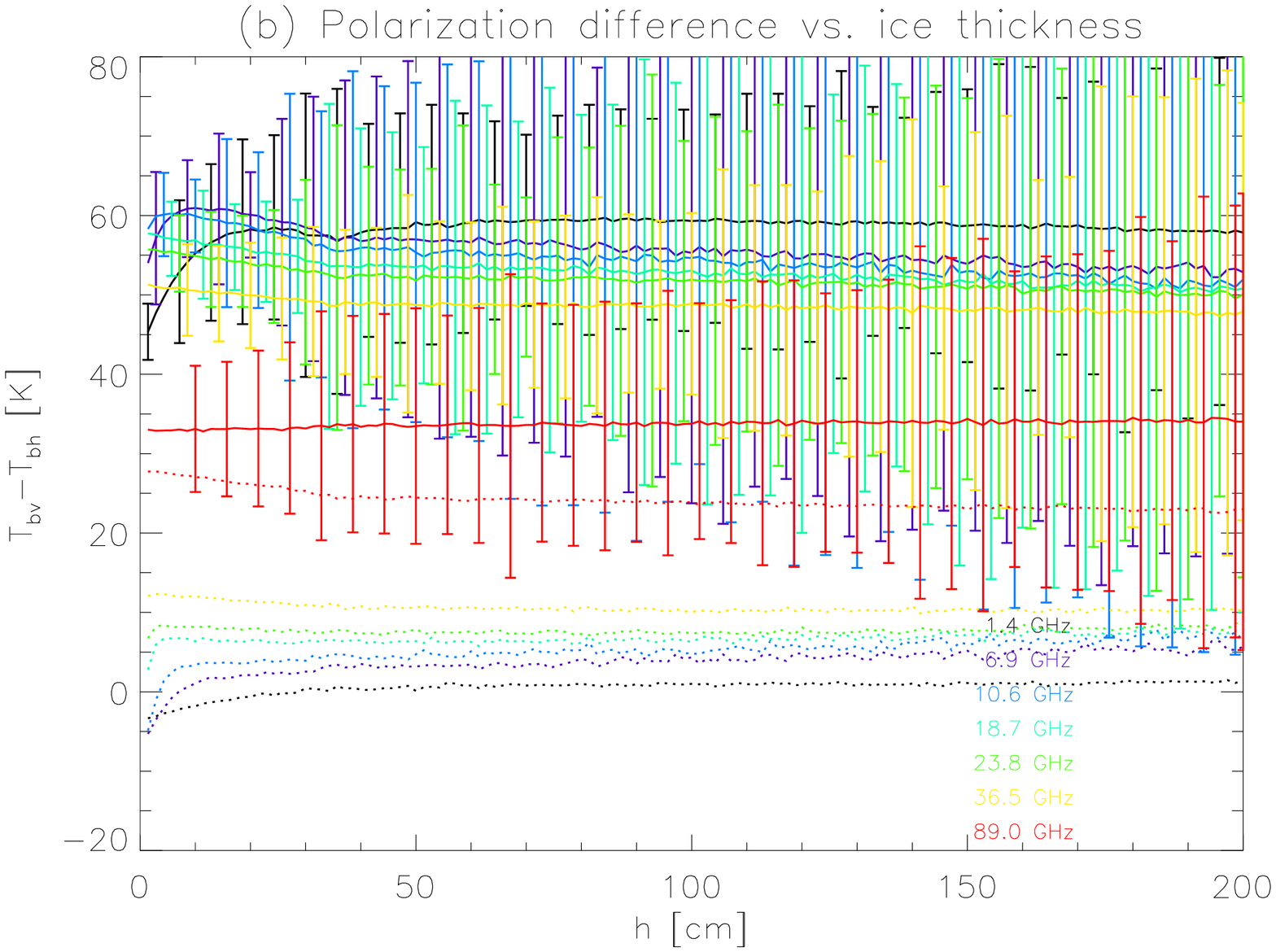}
\caption{Modelled brightness temperature as a function of ice thickness
using a single ice layer and constant temperature of 265K.
Dotted lines show scattering component.}
\label{tbvsh1}
\end{center}
\end{figure}

\begin{figure}
\begin{center}
\includegraphics[angle=90,width=0.9\textwidth]{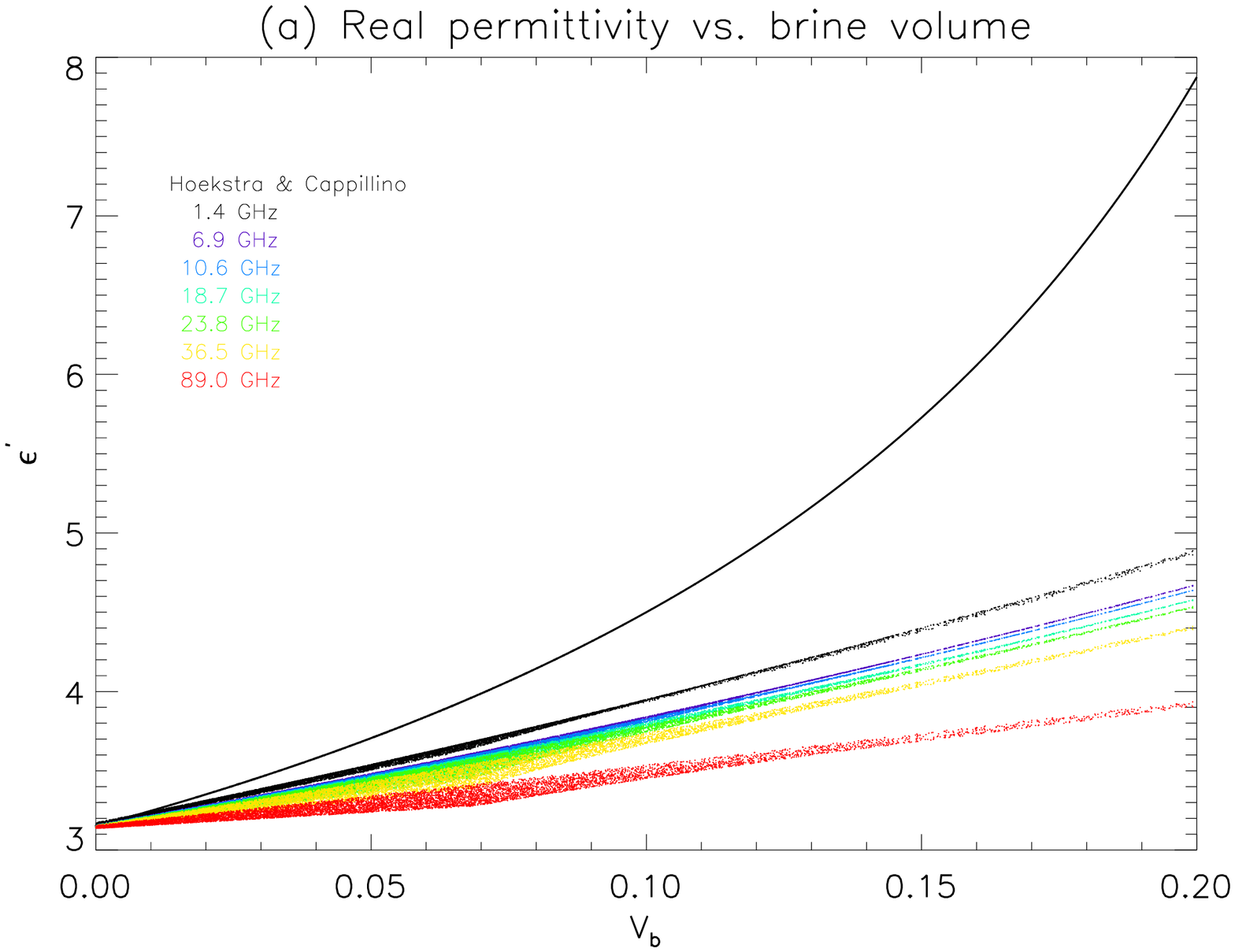}
\includegraphics[angle=90,width=0.9\textwidth]{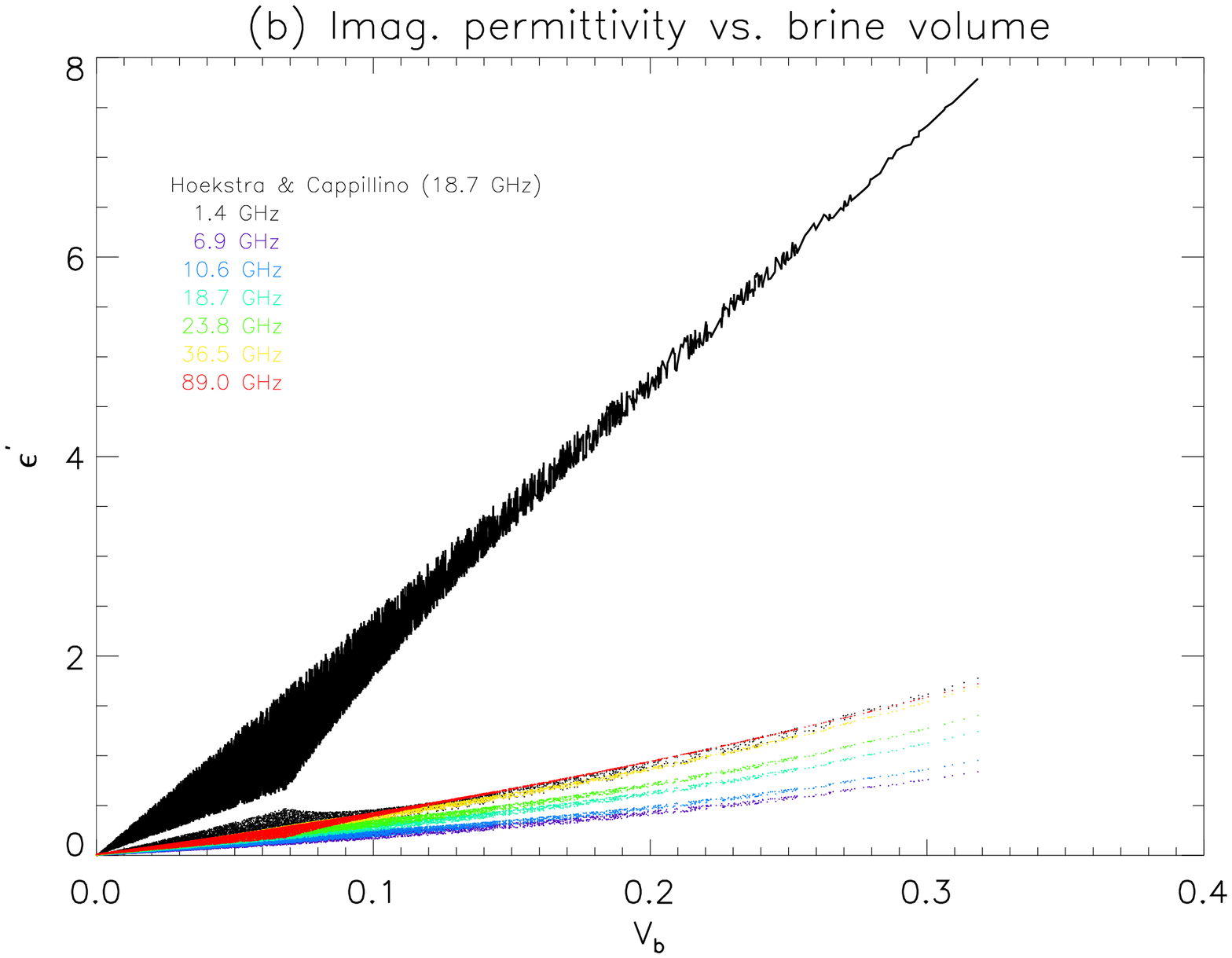}
\caption{Comparison of modelled effective permittivities using 
\citet{Hoekstra_Cappillino1971} and the semi-empirical model
described in the text.
The Hoekstra and Cappillino mixture model was used by
\citet{Naoki_etal2008} to model emissivity as a function of ice thickness.}
\label{epsvsvb}
\end{center}
\end{figure}

\begin{figure}
\begin{center}
\includegraphics[angle=90,width=0.9\textwidth]{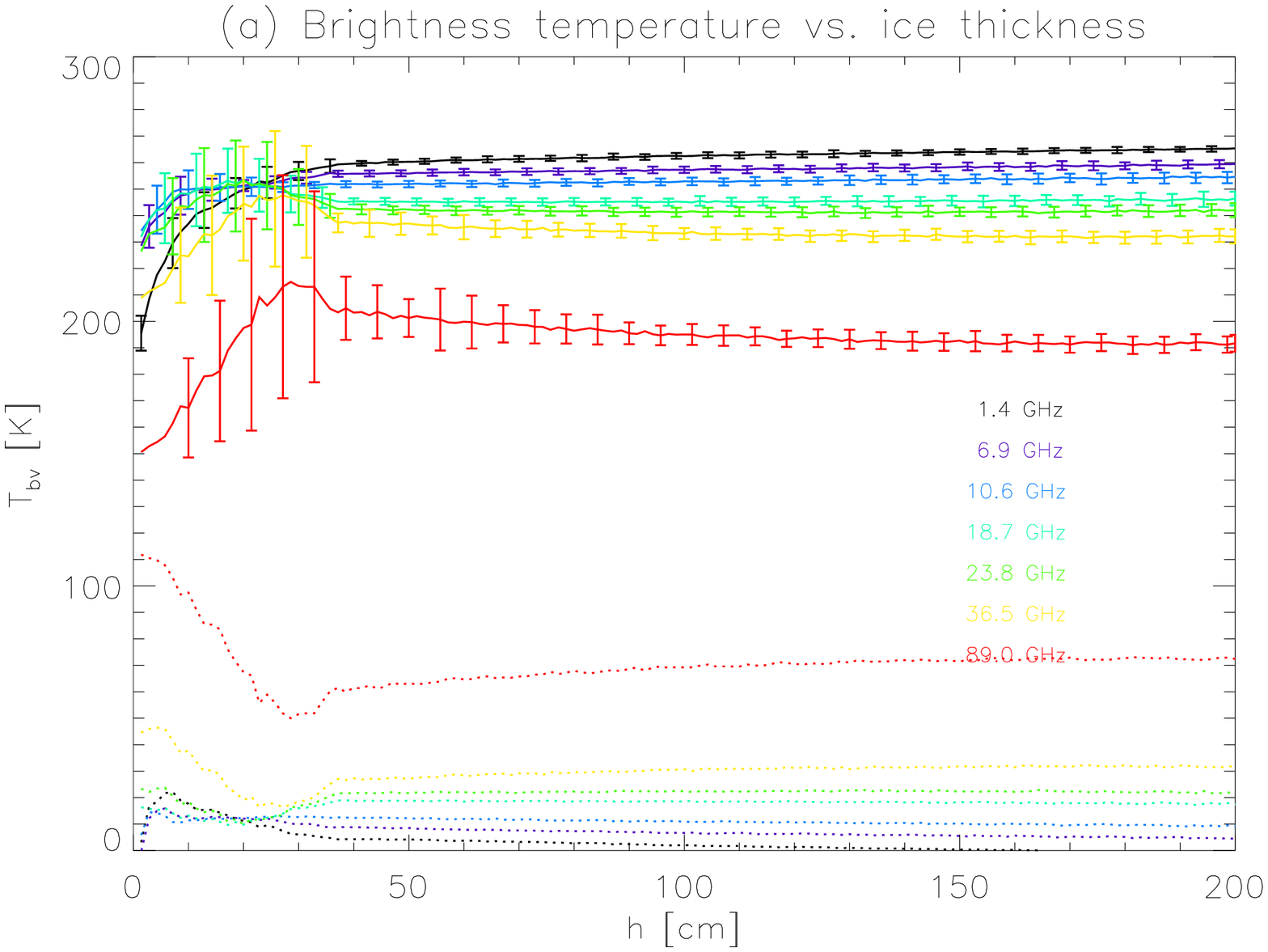}
\includegraphics[angle=90,width=0.9\textwidth]{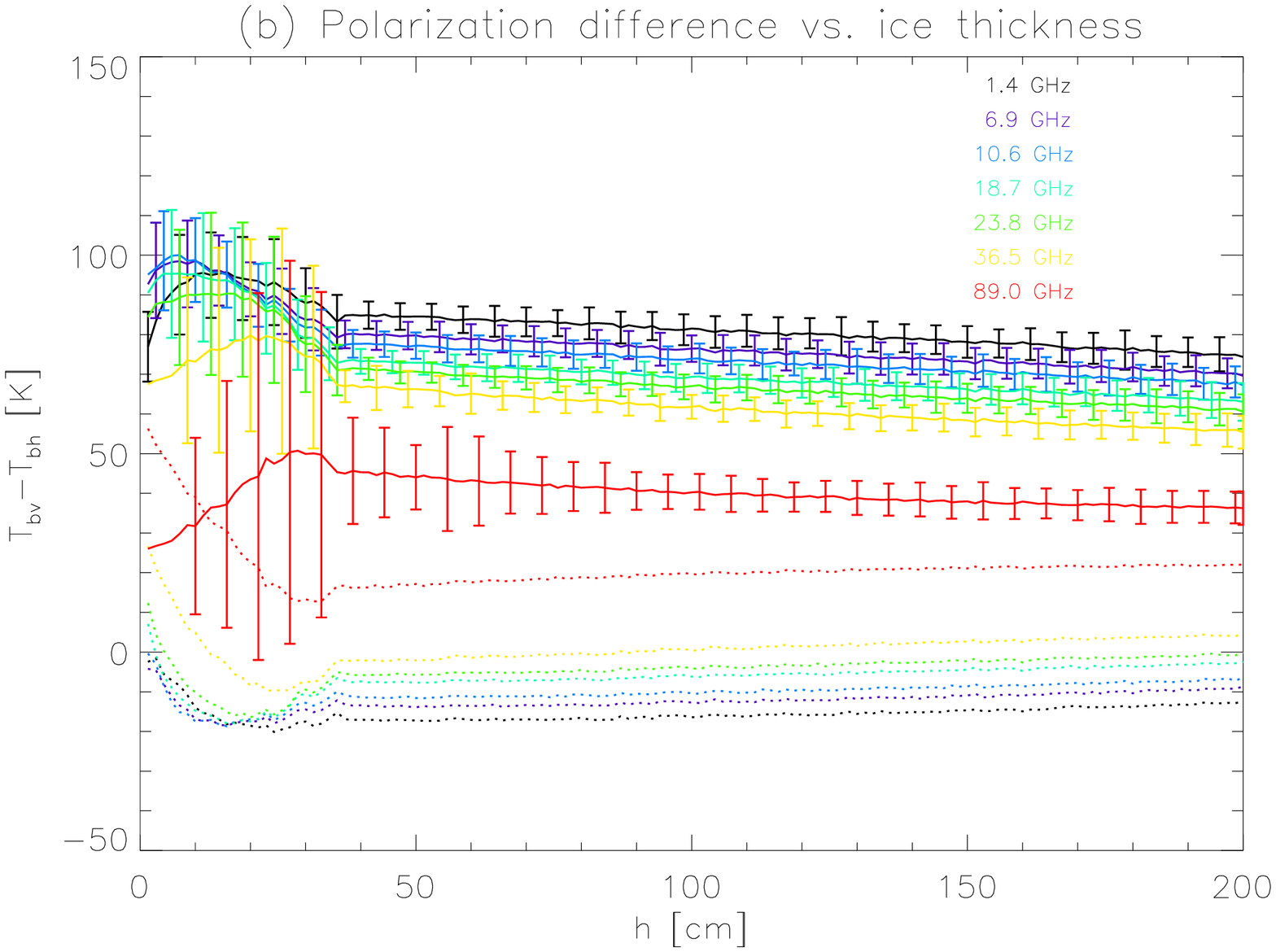}
\caption{Modelled brightness temperature as a function of ice thickness
using multiple (21) ice layers.  Salinity profile has an `S'-shape while
the temperature profiles is determined with a thermodynamic model.
Both are described in the text.
Dotted lines show scattering component.}
\label{tbvsh1a}
\end{center}
\end{figure}

Here the final $T_b$-thickness relations are presented.
The error bounds are rigorous in the sense that those quantities with known
error have been fully accounted for.
However,  there are many factors not accounted
for in the model such as the possible effects of ice ridging and anisotropy
of the brine inclusions.
Factors without known error, such as the correlation length of the brine inclusions,
have not been accounted for.

The dependence on ice thickness of vertically polarized brightness temperature
for the single (ice) layer model
 is shown in Figure \ref{tbvsh1} along with error bars.
Ice temperature was assumed to be 365 K.
What these results show is a modest and fairly rapid increase in emissivity
with thickness as the ice becomes progressively more opaque, thus less of the water
shows through.
Once the signal becomes ``saturated'' with only the ice emissions,
the curve remains relatively flat, with an emissivity of close to one.
The distance to saturation is controlled by the frequecy:
from Equation (8) in \citet{Mills_Heygster2011a},
the attenuation coefficient depends directly on the frequency.
For all but the bottom two or three frequencies, the water
layer hardly shows through even at the thinnest ice thickneses
resulting in an almost constant emissivity.

These results contrast with the those of \citet{Naoki_etal2008}
in which both modelled and measured ice emissivity was shown
to have a strong and slow increase with thickness.
The reason for the difference lies in the modelled effective
permittivies, which are shown in Figure \ref{epsvsvb}.
Naoki et al. use a mixture model that generates much larger
values at higher salinities.
It is this large real component that produces low emissities
at thinner ice thicknesses.
The changes in real permittivity will also produce higher
polarisation differences at thinner ice thickness
and this pattern can be seen weakly in Figure \ref{tbvsh1} (b).
(See \citet{SMOSIce_report} and \citet{Mills_Heygster2011a}
for an explanation.)
It is this decrease in polarization difference that is frequently used
to detect thin ice, as in \citet{Martin_etal2005}.

The $T_b$-thickness curves for the more sophisticated, multi-layer
model are shown in Figure \ref{tbvsh1a}.
21 layers were used in the model.
Note that this model shows a stronger increase/decrease
in brightness-temperature/polarization difference with
ice thickness, especially at higher frequencies.
This is due mainly to the scattering component, 
which is shown by the dotted lines.
Error bars for these curves are narrower because the 
random variations in effective permittivity in each
layer will tend to cancel.
Note that errors are higher for the polarization differences
(in both models) because they were calculated from the root-sum-square of
the errors in the horizontal and vertical.

\section{Summary and conclusions}

Radiative transfer-based emissivity models were run
for all frequencies of the Advanced Microwave Scanning
Radiometer on EOS (AMSR-E) instrument 
(6.925, 10.65, 18.7, 23.8, 36.5 and 89.0 GHz) as well
as 1.4 GHz in order to determine the relationship of
emissivity to sea ice thickness.
Because of ice growth processes, ice salinity
is normally inversely related
to ice thickness.
The satistical model of \citet{Cox_Weeks1974} 
was used to relate bulk salinity to ice thickness.
The emissivity model assumes plane parallel geometry;
two versions were employed: one with only one ice layer
(uniform properties throughout) and one with multiple
layers.  
For the single layer model, a constant temperature
of 265 K was used.
For the multi-layer model, the salinity profile was
modelled with the `S'-shaped profile from \citet{Eicken1992},
scaled to match the bulk salinity.
For the temperature profile, a thermodynamic model was
employed which assumed thermal equilibrium and
weather conditions equivalent to fall freeze-up.

Complex effective permittivities to feed to the RT
model were calculated from salinities and temperatures.
\citet{Vant_etal1978} derived measurement-based, statistical models
that linearly relate effective permittivity to brine
volume.
These models were only generated for frequencies between 0.1 and 4 GHz.
In order to extend the Vant models to higher frequencies,
 theoretical mixture models taken from
\citet{Sihvola_Kong1988} and based on the low
frequency limit were statistically adjusted
to the Vant models.

Error tolerances were calculated using a Monte Carlo
method based on the residuals supplied with both the
Cox and Weeks salinity-thickness relation and 
with the Vant effective permittivity models.

When scattering is neglected, the brightness temperature
is found to vary only slightly with 
ice thickness for both the single- and multi-layer models,
except at the lower frequecies, where brightness temperature
increases with ice thickness.
The latter effect is caused by the greater translucency of the ice
at lower frequencies: the water below shows through.
The reason little change is observed at higher frequencies
is because the effective permittivity models show only a weak
increase in real permittivity with brine volume.
Contrast this result to \citet{Naoki_etal2008}.
Note that both the real and imaginary permittivities are almost linearly
related to salinity.

Adding scattering produces much more variation with ice thickness of both
brightness temperature and polarization difference,
particularly in the multi-layer model.
Below about 30 cm, the brightness temperature is seen to increase
sharply with ice thickness, while above that it decreases
gently, especially at the higher frequencies.
This is in accordance with measurements where it is observed
that thin ice has both a lower brightness temperature
and higher polarization difference, while older, thicker
ice tends to be radiometrically cooler.
Meanwhile, the size of the scatterers
ensure that scattering is greater at higher frequencies.
It also makes sense that most of this difference is due to
scattering: thinner ice is more saline, hence it will have
both more and larger brine pockets.  It also tends to be
more granular, although this is not well accounted for in
the models.
For the multi-layer model, scattering in the horizontal
polarization is larger than that in the vertical,
a possibly questionable result.

One puzzling feature of the muliti-layer results is the low polarization
difference for very thin ice (below 20 cm), especially
at higher frequencies.  This is due to the scattering
component and does not accord well with measurements.
To correct this, a better understanding of the relationship
between scattering and ice physical properties
is necessary.

Error bounds were larger for the single-layer model
and reached as high as 50 K.
For the multi-layer model, bounds high
for thin ice thicknesses (again, as high as 50 K) but
tended to be lower.  Most were below 10 K, particularly
for lower frequencies and higher ice thicknesses.

Better ice emissivity models require better estimates
of ice effective permittivity, hence the design of a new,
combined model.
The correlation in wavelength of the Vant empirical
models with the Sihvola and Kong theoretical models
(based on the low-frequency limit) is likely significant,
but a fuller understanding would require a deeper investment
into the theory.
A similar statement is true for the scattering components
of the model in which the correlation length was
related to brine volume using a simple, ad hoc assumption.

Understanding ice growth processes would be another area
for future study since they are closely connected with
ice emissivity.  A simple ice growth model based on the
thermodynamic model for ice temperature was used to 
qualitatively confirm the relationship of bulk salinity
to thickness.
Ice cores from the Weddell Sea were similarly modelled
however the results are not shown here.

\bibliography{../final/sea_ice,../final/sft,../final/smos_final_extra,../final/smos_wp2.3a.bib,tstudy}

\end{document}